\documentclass[12pt]{iopart}
\begin{document}

\title[Apparent violation of thermodynamics second law]
{Apparent violation of thermodynamics second law
under Van der Waals gas expansion in vacuum}

\author{Yu.A.Koksharov}
\address{General Physics Department, Faculty of Physics,
M.V.Lomonosov Moscow State University, Moscow, Russia, 119992}
\ead{koksharov@nm.ru}
\submitto{EJP}

\begin{abstract}
Examining textbooks of thermodynamics we have not able to find
a proof of increasing of entropy under free adiabatic expansion in vacuum
based on the state equation of the Van der Waals gas.
This communication remedies the situation. During proving we faced
with an amusing example of apparent violation of thermodynamics second law. 
\end{abstract}

The paradoxes and mistakes are not rare in textbooks of thermodynamics [1,2]. 
Many of them are concerned with entropy and the second law of thermodynamics [1-3].
An analysis of thermodynamics paradoxes is instructive for students and can be  
usuful supplement to ordinary problems.

The second law is often deduced in classical thermodynamics as
a principle of increasing of entropy in irreversible adiabatic processes
[4].
For example, when a gas makes a free adiabatic expansion in vacuum
its entropy $S$ should increase.
This is true for any gas regardless of its state equation
due to the following general formula
[5]:
$$
{\partial S\over \partial V}\Big|_{U}={p\over T}>0,
$$
where $V$ is the system volume,
$U$ is the internal energy,
$T$ is the absolute temperature, $p$ is the pressure.

In case of an ideal gas the increase of entropy 
in the process of free adiabatic expansion in vacuum
can be shown plainly
using the first law of thermodynamics
$$
\delta Q=dU+\delta A,
\eqno(1)
$$
where
$\delta Q$ is the heat received by the gas from
its surroundings,
$\delta A$ is the work of external forces,
which surroundings exert on the gas;
the entropy definition
$$
dS={\delta Q\over T};
\eqno(2)
$$
and the equation of state of an ideal gas
$$
pV=\nu RT,
$$
where R is the gas constant, $\nu$ is the mole number.

Indeed, in that case any infinitesimal change of $S$ is equal to
$$
dS={1\over T}(dU+\delta A)=
{1\over T}(C_V dT + pdV)=
C_V{dT\over T}+\nu R{dV\over V},
\eqno(3)
$$
where $C_V$ is the isovolumic specific heat.

The integration of equation (3)  yields a change of
the entropy when the ideal gas goes from an initial state 1
to a final state 2:
$$
\Delta S=C_V\ln({T_2\over T_1})+\nu R\ln({V_2\over V_1})
\eqno(4)
$$

Note that though equations (2,3) are valid only for reversible processes,
equation (4) is applicable also to non-reversible processes,
if initial and final states are thermal equilibrium.

We consider the adiabatic process, and, therefore, $Q=0$.
Besides, the total system volume $V_2=const$, and we get $ A=0$.
Hence, $\Delta U=Q-A=0$  and for the ideal gas
$$
\Delta T={\Delta U\over C_V}=0 \rightarrow T_2=T_1 .
$$

Using equation (2) and taking into account $V_2>V_1$, we get
$$
\Delta S=\nu R\ln({V_2\over V_1})>0.
$$

A more complicate task is to develop a simple proof
of entropy increasing under adiabatic expansion in vacuum
in case of the Van der Waals gas.
The state equation of the Van der Waals gas is written usually as
[6]
$$
(p+{a\over V_\mu^2})(V_\mu-b)=RT,
\eqno(5)
$$
where $V_\mu={V\over \nu}$ is the molar volume, $a,b$
are Van der Waals constants.

Using equations (1,5) we get formulas
for changes of the entropy and the internal energy
of the Van der Waals gas [7]:
$$
\Delta S=C_{\mu,V}\ln({T_2\over T_1})+
R\ln({V_{\mu,2}-b\over V_{\mu,1}-b})
\eqno(6)
$$
$$
\Delta U=C_{\mu,V}(T_2 - T_1)+
a({1\over V_{\mu,1}} - {1\over V_{\mu,2}}),
\eqno(7)
$$
where $C_{\mu,V}$ is the isovolumic molar specific heat.

Since $\Delta U=0$, it follows from equation (5):
$$
T_2=T_1-{a(V_{\mu,2}-V_{\mu,1})\over C_{\mu,V}V_{\mu,1}V_{\mu,2}}
\eqno(8)
$$

Using $V_{\mu,2}>V_{\mu,1}$, from equation (8) we get $T_2<T_1$,
i.e., the Van der Waals gas always refrigerates
under free adiabatic expansion in vacuum.

Substituting equation (8) for $T_2$ in equation (6), we obtain:
$$
\Delta S=C_{\mu,V}\ln(1-{a(V_{\mu,2}-V_{\mu,1})
\over C_{\mu,V}V_{\mu,1}V_{\mu,2}T_1})+
R\ln({V_{\mu,2}-b\over V_{\mu,1}-b})
\eqno(9)
$$

Taking into account $V_{\mu,2}>V_{\mu,1}>b$, we conclude
that in equation (9) the second item is always positive, while
the first item is always negative.
One would think, that assuming
the values of $a,b,V_{\mu,1},V_{\mu,2}$ in equation (9) are fixed,
we could choose $T_1$ sufficiently small, so that
the sum (9) becomes negative (or equal to zero).
However, {\bf this conclusion conflicts with the second thermodynamics law}.
Probably, our assumption is not correct.
Let us prove that.

We begin with an inequality (see equation (6)):
$$
\Delta S=C_{V}\ln({T_2\over T_1})+
R\ln({V_{\mu,2}-b\over V_{\mu,1}-b})>0
\eqno(10)
$$
From equation (10) it follows that:
$$
({T_2\over T_1})({V_{\mu,2}-b\over V_{\mu,1}-b})^{R\over C_V}>1 \rightarrow
$$
$$
T_2 (V_{\mu,2}-b)^{R\over C_V}> T_1 (V_{\mu,1}-b)^{R\over C_V}.
\eqno(11)
$$

Substituting $T_2$ from equation (8) in equation (11), we get:
$$
T_1((V_{\mu,2}-b)^{R\over C_V} -(V_{\mu,1}-b)^{R\over C_V}) >
(V_{\mu,2}-b)^{R\over C_V} {a\over C_V}
{V_{\mu,2}-V_{\mu,1}\over V_{\mu,1}V_{\mu,2}}
\eqno(12)
$$

Assuming $V_{\mu,2}=V_{\mu,1}+dV$, where $dV<<V_{\mu,1}$,
from equation (12) we have:
$$
{RT\over C_V}(V_{\mu}-b)^{{R\over C_V}-1}dV >
(V_{\mu}-b)^{R\over C_V}{a\over C_V}{dV\over V_{\mu}^2}\rightarrow
$$
$$
RT > a {V_{\mu}-b\over V_{\mu}^2},
\eqno(13)
$$
where $V_{\mu}=V_{\mu,1}$, $T=T_1$.

However, from equation (5) it follows that
$$
a {V_{\mu}-b\over V_{\mu}^2}=RT-p(V_{\mu}-b)<RT,
\eqno(14)
$$
when $p>0$ and $V_{\mu}-b>0$.

It remains to note that equations (13) and (14) are equivalent.
Because all transformations from equation (10) to equation (13) are identical,
we conclude that while the Van der Waals state equation (5) is valid,
equation (10) is true also.

This completes the formal proof of increasing of the entropy of
the Wan der Vaals gas in the infinitesimal process of the
free adiabatic expansion in vacuum.
The process with finite change of the volume can be regarded as
an integral sum of corresponding infinitesimal processes.
Hence, any process of free adiabatic expansion in vacuum for
the Van der Waals gas is characterized by the entropy increase.
Presupposed violation of thermodynamics second law is apparent since 
we can not vary parameters in equation (9) independently of one another.

\end{document}